\documentclass[12pt]{article}
\usepackage{amsmath,amscd}
\usepackage{amsfonts}
\usepackage{amssymb}
\usepackage{graphicx,shortvrb}
\usepackage{epsfig}
\usepackage{newlfont}
\usepackage{hyperref}
\usepackage{xcolor}
\hypersetup{
    colorlinks,
    linkcolor={red!50!black},
    citecolor={blue!50!black},
    urlcolor={blue!80!black}
}

\def\susyN{N}
\def\Lag{\cal{L}}
\def\dil{\varphi}
\def\threeformsix{H}

\def\detT{\det \Tmat}
\def\Tmat{T}
\def\covD{ \cal D}
\def\poped{\Delta}
\def\popeg{g_0}
\def\popem{\mu}
\def\popeu{U}

\def\gprime{k_0}
\def\volthree{{\mathrm{vol}_3}}

\def\twoformthree{F}
\def\oneformthree{A}

\def\scman{{\cal M}}

\def\labelsotF{{\mathrm F}}
\def\labelsota{{++}}
\def\labelsotb{{+-}}
\def\labelsotc{{-+}}
\def\labelsotd{{--}}

\def\labelsofR{{\mathrm R}}
\def\labelsofD{{\mathrm D}}

\def\fcoord{{y}}
\def\etametric{{\eta}}
\def\idmat{{\mathbf{1}}}

\def\soegen{L}
\def\torgen{P}
\def\qgen{Q}
\def\dsofgen{L_{\labelsofD}}

\def\emat{{E}}

\def\embed{{\Theta}}
\def\ga{{g_1}}

\def\Lagg{{\tilde \Lag}}

\def\Gmet{G}
\def\Mmet{M}
\def\cospt{{\tilde{\cal P}}}
\def\cosvt{{\tilde{\cal V}}}
\def\scalars{\tilde\phi}
\def\cosrept{{\tilde{\cal S}}}

\def\oneformthreeg{{\oneformthree}}
\def\twoformthreeg{{\twoformthree}}

\def\ften{{\rho}}

\def\Apot{{\cal A}}
\def\scalmet{{\mathbb{G}}}

\def\Tten{{\cal T}}

\def\Ync{{Y}}

\def\sotagen{{L}_\labelsota}

\def\acronymforthethreeDsuperpotential{W}
\def\theletterwhichIwilluseinthefollowingforthescalarpotential{V}

\def\Pp{\mathbb{P}_+}

\def\Gam{{\Gamma}}

\def\fk{{f}}

\def\cosrep{{\cal S}}
\def\cosv{{\cal V}}
\def\escalars{{\chi}}
\def\cosp{e}
\def\cosq{\omega}

\def\isixa{\mu}
\def\isixb{\nu}
\def\isixc{\rho}

\def\ifoura{i}
\def\ifourb{j}
\def\ifourc{k}
\def\ifourd{l}
\def\ifoure{m}
\def\ifourf{n}
\def\ifourg{p}
\def\ifourh{q}

\def\ithreea{\mu}
\def\ithreeb{\nu}
\def\ithreec{\rho}

\def\ieighta{A}
\def\ieightb{B}
\def\ieightc{C}
\def\ieightd{D}

\def\ipm{{\ifoura_\pm}}
\def\jpm{{\ifourb_\pm}}

\def\ip{{\ifoura_+}}
\def\jp{{\ifourb_+}}
\def\im{{\ifoura_-}}
\def\jm{{\ifourb_-}}
\def\kp{{\ifourc_+}}
\def\lp{{\ifourd_+}}
\def\mp{{\ifoure_+}}
\def\np{{\ifourf_+}}
\def\pp{{\ifourg_+}}
\def\qp{{\ifourh_+}}
\def\km{{\ifourc_-}}
\def\lm{{\ifourd_-}}

\def\iscala{{\Lambda}}
\def\iscalb{{\Omega}}

\def\igena{{\cal A}}
\def\igenb{{\cal B}}
\def\igenc{{\cal M}}
\def\igend{{\cal N}}

\def\isota{{r}}

\newcommand{\ncproj}[1]{{\langle #1 \rangle}}
\newcommand{\rproj}[1]{{| #1 |}}

\def \Tr{\mathrm{Tr}}

\begin{document}

\rightline{}
\rightline{\tt }
\rightline{\today}
\vspace{1cm}

\begin{center}
{\Large{\bf A supersymmetric reduction on the three-sphere} }
\end{center}

\vskip 2mm

\noindent
{{Nihat Sadik Deger${}^1$, Henning Samtleben${}^2$, \"{O}zg\"{u}r Sar{\i}o\u{g}lu${}^3$,
Dieter Van den Bleeken${}^4$}}
\vskip .9cm

\begin{center}
{\small
{\it {${}^1$ Department of Mathematics,
Bo\u{g}azi\c{c}i University, \\
34342, Bebek, Istanbul, Turkey}}\\
sadik.deger@boun.edu.tr
\vskip 0.2cm
{\it {${}^2$ Universit\'e de Lyon, Laboratoire de Physique, UMR 5672, CNRS}}\\
{\it {\'Ecole Normale Sup\'erieure de Lyon}}\\
{\it {46, all\'ee d'Italie, F-69364 Lyon cedex 07, France}}\\
henning.samtleben@ens-lyon.fr
\vskip 0.2cm
{\it {${}^3$ Department of Physics,
Middle East Technical University, \\
06800, Ankara, Turkey}}\\
sarioglu@metu.edu.tr
\vskip 0.2cm
{\it {${}^4$ Department of Physics,
Bo\u{g}azi\c{c}i University, \\
34342, Bebek, Istanbul, Turkey}}\\
dieter.van@boun.edu.tr
}
\end{center}

\vskip 4mm

\abstract{We present the embedding of three-dimensional SO(4)$\ltimes\mathbb{R}^6$ gauged $\susyN=4$ supergravity with 
quaternionic target space SO(4,4)$/$(SO(4)$\times$SO(4)) into $D=6$, $\susyN=(1,0)$ supergravity coupled to a single chiral 
tensor multiplet through a consistent reduction on AdS$_3\times S^3$.}

\section{Introduction}
The subject of dimensional reduction continues to play a central role in string and supergravity theory, due to its phenomenological applications as well as the insights it provides into the structure of the theories and various dualities between them. Sphere reductions are especially interesting since they are prime examples where the problem of consistency\footnote{ A reduction is consistent if all solutions of the lower dimensional theory are also solutions of the higher dimensional one.} shows up and often can be addressed. 
Furthermore the positive curvature of the sphere sometimes can be balanced in the lower dimensional theory by an AdS vacuum. This makes sphere reductions useful in problems that rely on AdS/CFT duality conjecture. 

Known consistent sphere reductions include $S^7$ \cite{deWit:1986iy} 
and $S^4$ \cite{Nastase:1999cb,Nastase:1999kf}
compactifications of $D=11$ supergravity, and sectors of the $S^5$ reduction of Type IIB \cite{Cvetic:2000nc}.
They also include the $S^3$ and $S^4$ reductions of Type IIA supergravities \cite{Cvetic:2000ah}, which do not have 
AdS but domain-wall vacua.
These are examples of consistent embeddings of gauged supergravity theories 
with maximum supersymmetry in $D= 4, 5, 6, 7$. 
There are also examples with half supersymmetry 
\cite{Lu:1999bc, Lu:1999bw, Cvetic:1999au, Karndumri:2014pla}. 
For a general review and the reduction ansatz see \cite{Cvetic:2000dm, Cvetic:2000zu}. 
Finally, a somewhat peculiar consistent $S^2$ reduction of $D=6$, $N=(1,0)$ Einstein-Maxwell gauged supergravity was found in \cite{Gibbons:2003gp} which results in 
an $N=1$ (Minkowski)$_4$ vacuum.
In general, consistency of the reduction is unrelated to supersymmetry.

In this paper we revisit the consistent $S^3$ reduction of \cite{Cvetic:2000dm} for the case $D=6$ to $D=3$  which we observe to be an exception in the generic family. Integrating out the non-propagating two-form gauge potential,
the lower-dimensional Lagrangian gets an extra Chern-Simons contribution and
the scalar potential gains an additional term, thereby supporting an AdS vacuum. We show that actually the $D=3$ theory is the 
bosonic part of SO(4)$_\labelsofD\ltimes\mathbb{R}^6$ gauged $\susyN=4$ supergravity with quaternionic target space SO(4,4)$/$(SO(4)$\times$SO(4)). 
The $S^3$ reduction consistently embeds it into $D=6$ $\susyN=(1,0)$ supergravity coupled to a single chiral tensor multiplet. Since the bosonic parts of both supergravities are strongly constrained by supersymmetry which uniquely
fixes their coupling to fermions, we expect the relation between these two theories to hold for their fermionic parts as well, see also~\cite{Cvetic:2000dm} 
for a more detailed argument.

In section 2 we perform the 3-sphere reduction of the bosonic sector of $D=6$ $\susyN=(1,0)$ supergravity coupled to a single chiral tensor multiplet and shortly discuss the resulting $D=3$ theory. In the next section we then show in detail how this $D=3$ theory is the bosonic sector of a particular $N=4$ gauged supergravity. While doing so, we also show that the scalar potential of any  $D=3$, $N=4$ gauged supergravity whose target space is a single quaternionic manifold can be expressed in terms of a real superpotential, a result which can be of independent interest. In section 4 we compare some of the features of the reduction considered in this paper to various other known reductions from $D=6$ to $D=3$.

\section{The sphere reduction}\label{spheresec}
We start from the bosonic sector of $D=6$ $\susyN=(1,0)$ supergravity coupled to a single chiral tensor-multiplet (see e.g.\ \cite{Nishino:1986dc}):
\begin{equation}\label{Lag6}
\Lag_6=\sqrt{-g}\Big(R-\frac{1}{2}\partial_\isixa\dil\partial^\isixa\dil-\frac{1}{12}e^{-\sqrt{2}\dil}\threeformsix_{\isixa\isixb\isixc}\threeformsix^{\isixa\isixb\isixc}\Big)
\;.
\end{equation}
This theory falls into the class of Lagrangians considered in \cite{Cvetic:2000dm}, that allow for consistent $S^3$ reductions. Following that work we propose the following reduction ansatz to compactify the theory on the three-sphere:
\begin{eqnarray}
ds_6^2&=&(\detT^{\frac{1}{4}})\left(\poped^{\frac{1}{2}}ds_3^2+\popeg^{-2}\poped^{-\frac{1}{2}}\Tmat^{-1}_{\ifoura\ifourb}\covD\popem^\ifoura \covD\popem^\ifourb\right),\nonumber\\
\dil &=& \frac{1}{\sqrt{2}}\log\left(\poped^{-1}\detT^\frac{1}{2}\right)\label{Gansatz} , \\
\threeformsix &=&\gprime (\detT) \,\volthree-\frac{1}{6}\epsilon_{\ifoura\ifourb\ifourc\ifourd}\left(\popeg^{-2}\popeu\poped^{-2}\popem^\ifoura
\covD\popem^\ifourb\wedge\covD\popem^\ifourc\wedge\covD\popem^\ifourd\right.\nonumber\\
&& \left.+3\popeg^{-2}\poped^{-2}\covD\popem^\ifoura\wedge\covD\popem^\ifourb\wedge\covD\Tmat_{\ifourc\ifoure}
\Tmat_{\ifourd\ifourf}\popem^\ifoure\popem^\ifourf 
+3\popeg^{-1}\poped^{-1}\twoformthree^{\ifoura\ifourb}\wedge\covD\popem^\ifourc\Tmat_{\ifourd\ifoure}
\mu^\ifoure\right) ,\nonumber
\end{eqnarray}
where
\begin{eqnarray}
\popem^\ifoura\popem^\ifoura&=&1\,,\qquad \poped=\Tmat_{\ifoura\ifourb}\popem^\ifoura\popem^\ifourb\,,
\qquad\popeu=2\,\Tmat_{\ifoura\ifourc}\Tmat_{\ifourb\ifourc}\popem^\ifoura\popem^\ifourb
-\poped\Tmat_{\ifoura\ifoura}\,,\nonumber\\
\covD\popem^\ifoura&=&d\popem^\ifoura+\popeg\oneformthree^{\ifoura\ifourb}\popem^\ifourb\,,\qquad
\covD\Tmat_{\ifoura\ifourb}=d\Tmat_{\ifoura\ifourb}+\popeg\oneformthree^{\ifoura\ifourc}
\Tmat^{\ifourc\ifourb}+\popeg\oneformthree^{\ifourb\ifourc}
\Tmat^{\ifourc\ifoura}\,,\\
\twoformthree^{\ifoura\ifourb}&=&d\oneformthree^{\ifoura\ifourb}+\popeg\oneformthree^{\ifoura\ifourc}\wedge \oneformthree^{\ifourc\ifourb}\,, \qquad\qquad \ifoura,\ifourb=1,\ldots,4\,.\nonumber
\end{eqnarray}

The ansatz is essentially identical to the one made in \cite{Cvetic:2000dm}, except that in this special case of reduction to three dimensions the external part of the field strength~$\threeformsix$, i.e.\ the first term in the ansatz \eqref{Gansatz}, becomes non-dynamical (it corresponds
to the field strength of a two-form gauge potential in three dimensions) and is completely fixed up to the overall constant $\gprime$.
The reduction of the $D=6$ equations of motion under this ansatz can be performed as in \cite{Cvetic:2000dm}. In this case they become effective $D=3$ equations that follow from the Lagrangian
\begin{equation}\label{3dtheory}
\Lag_3=\sqrt{-g}\left(R-\frac{1}{4}\,\Tmat^{-1}_{\ifoura\ifourb}\Tmat^{-1}_{\ifourc\ifourd}\covD_\ithreea \Tmat_{\ifourb\ifourc}\covD^\ithreea\Tmat_{\ifourd\ifoura}-\frac{1}{8}\,\Tmat^{-1}_{\ifoura\ifourc}\Tmat^{-1}_{\ifourb\ifourd}\twoformthree^{\ifoura\ifourb}_{\ithreea\ithreeb}
\twoformthree^{\ifourc\ifourd\, {\ithreea\ithreeb}}-\theletterwhichIwilluseinthefollowingforthescalarpotential\right)+\Lag_{\mathrm{CS}}
\;.
\end{equation} 
This Lagrangian is a slight exception in the general family considered in \cite{Cvetic:2000dm}. The 3-form field strength that is generically present is non-dynamical in three dimensions and disappears. Instead the scalar potential gains an additional term proportional to $\gprime$:
\begin{equation}\label{potred}
\theletterwhichIwilluseinthefollowingforthescalarpotential
=\frac{1}{2}\left(\gprime^2\detT+2\popeg^2\Tmat_{\ifoura\ifourb}\Tmat_{\ifoura\ifourb}-\popeg^2(\Tmat_{\ifoura\ifoura})^2\right)
\;.
\end{equation}
Additionally, in this particular case there is also a Chern-Simons term: 
\begin{equation}\label{CSterm}
\Lag_{\mathrm{CS}}=-\frac{1}{8}\,\gprime \epsilon_{\ifoura\ifourb\ifourc\ifourd}\varepsilon^{\ithreea\ithreeb\ithreec}
\oneformthree^{\ifoura\ifourb}_\ithreea\,\Big(\partial_\ithreeb \oneformthree^{\ifourc\ifourd}_{\ithreec}+\frac{2}{3}\,\popeg\oneformthree^{\ifourc\ifoure}_\ithreeb\oneformthree^{\ifoure\ifourd}_\ithreec\Big).
\end{equation}
It is interesting to note that the Chern-Simons term \eqref{CSterm} is not the standard SO(4) Chern-Simons term, but rather a sum of two SO(3) Chern-Simons terms of opposite level.
In the three-dimensional language, both the extra contribution to the potential and the Chern-Simons term arise from eliminating the non-physical three-form field strength from the generic Lagrangian of \cite{Cvetic:2000dm} by its equations of motion.

Let us point out that the extra term in the potential (\ref{potred}) stabilizes it at the origin and allows it to support an AdS$_3$ ground state, unlike the
generic reductions in \cite{Cvetic:2000dm, Cvetic:2000zu}. It is instructive to isolate the $\mathrm{det}\,\Tmat\equiv e^{4\sigma}$ factor which is a singlet under the 
gauge group as
\begin{eqnarray}
\theletterwhichIwilluseinthefollowingforthescalarpotential
=\frac{1}{2}\left(\gprime^2e^{4\sigma}+\popeg^2 e^{2\sigma}\left(2\, \Tr\,\hat\Tmat^2-(\Tr\,\hat\Tmat)^2\right) \right)
\;,
\label{pot3}
\end{eqnarray}
with a matrix $\hat\Tmat$ of unit determinant.
This shows that the relative coefficient $g_0^2/k_0^2$ can simply be absorbed into a shift in $\sigma$.
We may choose to set $k_0=2g_0$, in which case the scalar potential has its extremal point at the origin $\sigma=0$,
$\hat{T}=\mathbb{I}_4$\,, which corresponds to the AdS$_3\times S^3$ vacuum in $D=6$, as can be seen from \eqref{Gansatz}.
Around this origin, which is a supersymmetric stationary point, the 9 scalars from $\hat\Tmat$ come 
with zero mass, whereas the dilaton $\sigma$ has mass corresponding to conformal dimension $\Delta=4$\,.

The particular features that set apart the 3-sphere reduction to three dimensions in the general class considered in \cite{Cvetic:2000dm} turn out to be nothing but consequences of underlying supersymmetry. Actually, the resulting Lagrangian \eqref{3dtheory} precisely corresponds to the bosonic sector of a $D=3$ $\susyN=4$ gauged supergravity, as we will show in the rest of this paper. In particular, this embedding of the Lagrangian (\ref{3dtheory}) into the general class of $N=4$ gauged supergravities
allows its fermionic couplings to be read off from~\cite{deWit:2003ja}.

\section{The $D=3$ gauged supergravity}\label{gssection}
Any three-dimensional $N=4$ supergravity can be formulated as a gauged linear sigma-model coupled to gravity and Chern-Simons gauge fields \cite{deWit:2003ja} and is uniquely determined by the choice of a quaternionic manifold as the scalar target space and an embedding tensor describing the gauge-group. In the case relevant for the above reduction, the quaternionic manifold will be the coset manifold
\begin{equation}
\scman=\frac{\mathrm{SO(4,4)}}{\mathrm{SO(4)\times SO(4)}}
\label{so4444}
\end{equation}
and the Chern-Simons gauge-group will be the semi-direct product SO(4)$\ltimes\mathbb{R}^6$. For such a semi-direct gauging there is an alternative formulation of the theory where some of the scalars are dualised into Yang-Mills gauge fields \cite{Nicolai:2003bp}. It is in this second formulation that the bosonic part of the theory becomes identical to \eqref{3dtheory}, as we will now show by explicitly constructing it from its definition.

\subsection{Symmetries and gauging}
The global symmetry of the ungauged $N=4$ theory with target space (\ref{so4444}) is given by 
SO(4,4)$\times$SO(3)$_{\labelsotF}$ symmetry with maximally non-compact subgroup 
\begin{equation}
{\mathrm{SO}}(3)_\labelsota
\times{\mathrm{SO}}(3)_\labelsotb
\times{\mathrm{SO}}(3)_\labelsotc\times{\mathrm{SO}}(3)_\labelsotd\times{\mathrm{SO}}(3)_\labelsotF
\;.
\end{equation}
The R-symmetry is\footnote{ One can choose any of the first four SO(3) factors to be the one associated to the R-symmetry, these choices are equivalent up to a discrete automorphism of the symmetry group and leave the physics invariant. We choose the first factor.} 
SO(4)$_\labelsofR=$SO(3)$_\labelsota\times$SO(3)$_\labelsotF$, where the second factor acts exclusively on the fermions. In the rest of this paper the SO(3)$_\labelsotF$ will play no further role (as it will remain ungauged) and so we will only focus on the SO(4,4) symmetry that acts on the scalar coset space. It will be convenient to describe its algebra using light-cone coordinates. We introduce the eight coordinates $\fcoord^\ieighta$, $\ieighta=1,\ldots, 8$,  so that the SO(4,4) invariant
metric $\etametric$ has the form
\begin{equation}
\etametric=\begin{pmatrix}
0&\idmat\\
\idmat&0
\end{pmatrix}\,.
\end{equation}
It will be useful to split these 8 directions in the first and last 4, so we introduce the notation 
\begin{equation}
\fcoord^\ip\equiv\fcoord^\ifoura\,,\quad \fcoord^{\im}\equiv\fcoord^{\ifoura+4}\,,\qquad \ifoura=1,\ldots,4.
\end{equation} For example, in this notation the metric $\eta$ has the following components:
\begin{equation}
\etametric_{\ipm\jpm}=0\,,\qquad \eta_{\ip\jm}=\eta_{\im\jp}=\delta_{\ifoura\ifourb} .
\end{equation} 
The generators $\soegen^{\ieighta\ieightb}$ of so(4,4) then split as:
\begin{eqnarray}
\torgen_\pm^{\ifoura\ifourb}\equiv \soegen^{\ipm\jpm}\,,\qquad \qgen^{\ifoura\ifourb}\equiv \soegen^{\ip\jm}-\soegen^{\im\jp}\,,\qquad \dsofgen^{\ifoura\ifourb}\equiv \soegen^{\ip\jm}+\soegen^{\im\jp} .
\end{eqnarray}
Here $\dsofgen^{\ifoura\ifourb}$ generates the diagonal SO(4)$_{\labelsofD}$ subgroup and all $\ifoura,\ifourb$ indices transform under the fundamental representation of this subgroup. The $\qgen^{\ifoura\ifourb}$ generators extend the SO(4)$_\labelsofD$ to a GL(4) subgroup. The $\torgen_\pm^{\ifoura\ifourb}$ form two sets of commuting nilpotent matrices and generate two $\mathbb{R}^6$ subgroups. The generators split as follows into compact and non-compact:
\begin{equation}
\torgen_+^{\ifoura\ifourb}+\torgen_-^{\ifoura\ifourb}\,,\quad \dsofgen^{\ifoura\ifourb}\quad \mbox{compact,}\qquad
\torgen_+^{\ifoura\ifourb}-\torgen_-^{\ifoura\ifourb}\,,\quad \qgen^{\ifoura\ifourb}\quad \mbox{non-compact.}
\end{equation}
It will be useful later to introduce a projection onto the non-compact part. As there are exactly sixteen such generators they can be labelled with a pair of $\ifoura\ifourb$ indices which we will write as $\ncproj{\ifoura\ifourb}$ to distinguish them. More precisely, for an adjoint so(4,4) valued tensor $\ften_{\ieighta\ieightb}$ we define\footnote{ Note that we use a very specific normalization for the non-compact generators. This normalization is directly related to the normalization of the target space metric and hence the kinetic term for the scalars, see section \ref{kinsec} and appendix \ref{norm}.}
\begin{eqnarray}
\ften_\ncproj{\ifoura\ifourb}&\equiv &\ften_{\ip\jp}-\ften_{\im\jm}+\ften_{\ip\jm}-\ften_{\im\jp}\,,\\
&\Rightarrow & \ften_{\ieighta\ieightb}\soegen^{\ieighta\ieightb}=\frac{1}{2}\,\ften_\ncproj{\ifoura\ifourb}\left(
\torgen_+^{\ifoura\ifourb}-\torgen_-^{\ifoura\ifourb}+\qgen^{\ifoura\ifourb}\right)+\mbox{compact}\,.
\end{eqnarray}

To perform computations it is useful to use an explicit matrix representations of these generators,
starting from 
  \begin{equation}
\left(\soegen^{\ieighta\ieightb}\right)_{\ieightc\ieightd}=-\delta^{\ieighta\ieightc}\eta^{\ieightb\ieightd}
+\delta^{\ieighta\ieightd}\eta^{\ieightb\ieightc}\,.
\end{equation}
Using the notation $\left(\emat^{\ifoura\ifourb}\right)_{\ifourc\ifourd}=\delta^{\ifoura\ifourc}\delta^{\ifourb\ifourd}$ one finds
\begin{eqnarray}
\torgen_+^{\ifoura\ifourb}&=&\begin{pmatrix} 0& -\emat^{\ifoura\ifourb}+\emat^{\ifourb\ifoura}\\ 0 & 0 \end{pmatrix}\,,\label{explgensa}\\
\torgen_-^{\ifoura\ifourb}&=&\begin{pmatrix} 0& 0\\ -\emat^{\ifoura\ifourb}+\emat^{\ifourb\ifoura} & 0 \end{pmatrix}\,,\\
\dsofgen^{\ifoura\ifourb}&=&\begin{pmatrix} -\emat^{\ifoura\ifourb}+\emat^{\ifourb\ifoura} & 0\\0&  -\emat^{\ifoura\ifourb}+\emat^{\ifourb\ifoura}  \end{pmatrix}\,,\\
\qgen^{\ifoura\ifourb}&=&\begin{pmatrix}-\emat^{\ifoura\ifourb}-\emat^{\ifourb\ifoura} & 0\\ 0&\emat^{\ifoura\ifourb}+\emat^{\ifourb\ifoura}  \end{pmatrix}
\;.\label{explgensz}
\end{eqnarray}

Now that we have specified the symmetry generators and indicated all the subgroups of interest in detail, we are ready to provide the last remaining piece of data, the embedding tensor $\embed_{\ieighta\ieightb,\ieightc\ieightd}$\,, which is valued in the symmetric product of the SO(4,4) adjoint representation and defines the gauge group generators according to
 \begin{equation}
 X_{AB} \equiv \Theta_{AB,CD}\,L^{CD}
\;.
\end{equation}
The gauging we will perform is specified by choosing the following non-zero components (up to symmetries in the indices), using the split $\ieighta=(\ip,\im)$ introduced above:
\begin{equation}\label{embed}
\embed_{\ip\jp,\kp\lp}=\frac{\gprime}{2} \epsilon_{\ifoura\ifourb\ifourc\ifourd}\,,\qquad \embed_{\ip\jp,\kp\lm}=-\frac{\popeg}{2}\left(\delta_{\ifoura\ifourc}\delta_{\ifourb\ifourd}-\delta_{\ifourb\ifourc}\delta_{\ifoura\ifourd}\right)\,. 
\end{equation}
Clearly this embedding tensor projects out the generators $\torgen_-^{\ifoura\ifourb}$. Since the component proportional to $\ga$ is automatically anti-symmetric in $\ifourc$ and $\ifourd$ it also projects out the generators $\qgen^{\ifoura\ifourb}$. In summary we are gauging the subgroup generated by the $\torgen_+^{\ifoura\ifourb}$ and $\dsofgen^{\ifoura\ifourb}$, which is SO(4)$_\labelsofD\ltimes\mathbb{R}^6$. Not all embedding tensors lead to consistent supersymmetric gaugings, but we will see that \eqref{embed} does in section \ref{potsec}.

\subsection{Lagrangian}
The specification of the symmetry algebra and the subalgebra that will be gauged provides all the ingredients that are needed to explicitly construct the Lagrangian of the respective $\susyN=4$  gauged supergravity. The recipe for constructing the Lagrangian of the theory in its Yang-Mills form can be found in \cite{Nicolai:2003bp} and leads to
\begin{eqnarray}
\Lagg_3&=&\sqrt{-g}\left(R-g^{\ithreea\ithreeb}\Gmet_{\ifoura\ifourb, \ifourc\ifourd}\cospt^{\ifoura\ifourb}_\ithreea\cospt^{\ifourc\ifourd}_\ithreeb-\frac{1}{8}\Mmet_{\ifoura\ifourb,\ifourc\ifourd}\twoformthreeg^{\ifoura\ifourb\, \ithreea\ithreeb}\twoformthreeg^{\ifourc\ifourd}_{\ithreea\ithreeb}-\theletterwhichIwilluseinthefollowingforthescalarpotential\right)\nonumber\\&&+\frac{1}{2}\varepsilon^{\ithreea\ithreeb\ithreec}\Mmet_{\ifoura\ifourb,\ifourc\ifourd}\cosvt^{\ip\jp}{}_{\ncproj{\ifoure\ifourf}}\twoformthreeg^{\ifourc\ifourd}_{\ithreea\ithreeb}\cospt^{\ifoure\ifourf}_\ithreec
+\Lagg_{\mathrm{CS}}
\;,
\label{3dgauged}
\end{eqnarray}
whose different terms we describe in the following.

\subsubsection{Kinetic terms}\label{kinsec}
The objects appearing in the kinetic term of (\ref{3dgauged}) are related to the scalar coset space:
\begin{eqnarray}
\cosvt^{\ieighta\ieightb}{}_{\ieightc\ieightd}\soegen^{\ieightc\ieightd}&=&\cosrept^{-1}\soegen^{\ieighta\ieightb}\cosrept\,,\qquad \cosrept=e^{\scalars_{\ifoura\ifourb}\qgen^{\ifoura\ifourb}}\,,\label{defs}\\
\Gmet_{\ifoura\ifourb, \ifourc\ifourd}&=&\delta_{\ifoura\ifourc}\delta_{\ifourb\ifourd}-\cosvt^{\mp\np}{}_{\ncproj{\ifoura\ifourb}}\Mmet_{\ifoure\ifourf,\ifourg\ifourh}\cosvt^{\pp\qp}{}_{\ncproj{\ifourc\ifourd}}\,,\nonumber\\
\Mmet_{\ifoura\ifourb,\ifourc\ifourd}&=&\left(\cosvt^{\ip\jp}{}_{\ncproj{\ifoure\ifourf}}\cosvt^{\kp\lp}{}_{\ncproj{\ifoure\ifourf}}\right)^{-1}\,,\nonumber\\
\cospt_\ithreea^{\ifoura\ifourb}&=&\left(\cosrept^{-1}\covD_\mu\cosrept\right)^{\ncproj{\ifoura\ifourb}}\,,\nonumber\\
\covD_\ithreea&=&\partial_\ithreea+\frac{1}{2}\Theta_{\ip\jp,\kp\lm}\oneformthreeg^{\ifoura\ifourb}_\ithreea\dsofgen^{\ifourc\ifourd}
\;.
\nonumber
\end{eqnarray}

The coset representative $\cosrept$ parameterizes the coset space
\begin{equation}
\frac{\mathrm{GL}(4)^+}{\mathrm{SO}(4)_\labelsofD}\subset\frac{\mathrm{SO}(4,4)}{\mathrm{SO}(4)\times\mathrm{SO}(4)}\,.
\end{equation} It provides the target space for the subset of scalars that are not dualized into dynamic vector fields \cite{Nicolai:2003bp}. 
In the representation of (\ref{explgensz}), we can parametrize the coset representative $\cosrept$ in terms of a positive definite $4\times4$ symmetric matrix $\Tmat_{\ifoura\ifourb}$ as follows
\begin{equation}\label{cosetpar}
\cosrept=e^{\scalars_{\ifoura\ifourb}\qgen^{\ifoura\ifourb}}=\begin{pmatrix}
\sqrt{\Tmat^{-1}}&0\\0&\sqrt{\Tmat}
\end{pmatrix}
\Leftrightarrow \Tmat=e^{4\scalars}
\;.
\end{equation}
One can then compute the objects (\ref{defs}) in this parametrization (we only spell out the non-zero components):
\begin{eqnarray}
\cospt_\ithreea^{\ifoura\ifourb}&=&\frac{1}{2}\left(\sqrt{\Tmat^{-1}}(\covD_\ithreea \Tmat)\sqrt{\Tmat^{-1}}\right)^{\ifoura\ifourb}\,,\nonumber\\
\covD_\ithreea \Tmat^{\ifoura\ifourb}&=&\partial_\ithreea \Tmat^{\ifoura\ifourb}+\popeg\left(\oneformthreeg_\ithreea^{\ifoura\ifourc}\Tmat^{\ifourc\ifourb}
+\oneformthreeg_\ithreea^{\ifourb\ifourc}\Tmat^{\ifourc\ifoura}\right)\,,\nonumber \\
\cosvt^{\ip\jp}{}_{\kp\lp}&=&\big(\sqrt{\Tmat}\big)_{\ifourc[\ifoura}
\big(\sqrt{\Tmat}\big)_{\ifourb]\ifourd}\,,\nonumber\\
\cosvt^{\im\jm}{}_{\km\lm}&=&\big(\sqrt{\Tmat^{-1}}\big)_{\ifourc[\ifoura}
\big(\sqrt{\Tmat^{-1}}\big)_{\ifourb]\ifourd}\,,\label{cosstuff}\\
\cosvt^{\ip\jm}{}_{\kp\lm}&=&\big(\sqrt{\Tmat}\big)_{\ifourc\ifoura}
\big(\sqrt{\Tmat^{-1}}\big)_{\ifourb\ifourd}\,,\nonumber\\
\Mmet_{\ifoura\ifourb,\ifourc\ifourd}&=&\Tmat^{-1}_{\ifoura[\ifourc}\Tmat^{-1}_{\ifourd]\ifourb}\,,\nonumber\\
\Gmet_{\ifoura\ifourb,\ifourc\ifourd}&=&\delta_{\ifoura(\ifourc}\delta_{\ifourd)\ifourb}\;.\nonumber
\end{eqnarray}
It follows that the sum of the kinetic terms in \eqref{3dgauged} is given by
\begin{eqnarray}
\sqrt{-g^{-1}}\Lagg_{\mathrm{Kin.}}&=&R-g^{\ithreea\ithreeb}\Gmet_{\ifoura\ifourb, \ifourc\ifourd}\cospt^{\ifoura\ifourb}_\ithreea\cospt^{\ifourc\ifourd}_\ithreeb-\frac{1}{8}\Mmet_{\ifoura\ifourb,\ifourc\ifourd}\twoformthreeg^{\ifoura\ifourb\, \ithreea\ithreeb}\twoformthreeg^{\ifourc\ifourd}_{\ithreea\ithreeb}\label{kin}\\
&=&R-\frac{1}{4}\Tmat^{-1}_{\ifoura\ifourb}\Tmat^{-1}_{\ifourc\ifourd}\covD_\ithreea
\Tmat_{\ifourb\ifourc}\covD^\ithreea
\Tmat_{\ifourd\ifoura}-\frac{1}{8}\Tmat^{-1}_{\ifoura[\ifourc}\Tmat^{-1}_{\ifourd]\ifourb}\twoformthreeg^{\ifoura\ifourb\, \ithreea\ithreeb}\twoformthreeg^{\ifourc\ifourd}_{\ithreea\ithreeb}\;.\nonumber
\end{eqnarray}
Note that this perfectly matches with the kinetic terms in \eqref{3dtheory}. Not only is the functional form identical, which is due to the underlying group theory, but also the relative normalization of the scalar kinetic term to the gravitational one coincides with \eqref{3dtheory}. This normalization is non-trivial and is directly linked to supersymmetry. The scale of the scalar coset geometry is fixed by the requirement that the K\"ahler forms $\fk^{\ifoura\ifourb}$ of its quaternionic structure are related to the curvature of the SO(3)$_\labelsota\subset$SO(4)$_\labelsofR$ connection $\omega^{\ifoura\ifourb}_\labelsota$ in a specific way  \cite{deWit:2003ja}:
\begin{equation}
d\cosq^{\ifoura\ifourb}_\labelsota+
\cosq^{\ifoura\ifourc}_\labelsota\wedge\cosq^{\ifourc\ifourb}_\labelsota=\frac{1}{2}\fk^{\ifoura\ifourb}\,.\label{normrel}
\end{equation}
We show this relation for the theory at hand in appendix \ref{norm}.

\subsubsection{Topological terms}
There are two topological terms in the Lagrangian \eqref{3dgauged}. The first vanishes in our model:
\begin{equation}
\varepsilon^{\ithreea\ithreeb\ithreec}\Mmet_{\ifoura\ifourb,\ifourc\ifourd}\cosvt^{\ip\jp}{}_{\ncproj{\ifoure\ifourf}}\twoformthreeg^{\ifourc\ifourd}_{\ithreea\ithreeb}
\cospt^{\ifoure\ifourf}_\ithreec=0
\;.
\end{equation}
This follows from the fact that $\cospt^{\ifourc\ifourd}_\ithreec$ is symmetric whereas $\cosvt^{\ip\jp}{}_{\ncproj{\ifourc\ifourd}}$ is anti-symmetric in ${\ifourc\ifourd}$ as can be seen from (\ref{cosstuff}). 
The Chern-Simons term is given by \cite{deWit:2003ja}
\begin{eqnarray}
\Lagg_{\mathrm{CS}}&=&-\frac{\gprime}{8}\varepsilon^{\ithreea\ithreeb\ithreec}\epsilon_{\ifoura\ifourb\ifourc\ifourd}
\oneformthree_\ithreea^{\ifoura\ifourb}\left(\partial_\ithreeb\oneformthree_\ithreec^{\ifourc\ifourd}
+\frac{2}{3}\popeg\oneformthree_\ithreea^{\ifourc\ifoure}\oneformthree_\ithreeb^{\ifoure\ifourd}\right)
\;.\label{CSg}
\end{eqnarray}
Note that the appearance of $\epsilon_{\ifoura\ifourb\ifourc\ifourd}$, or equivalently opposite levels for the two SO(3) factors, is directly related to the choice of embedding tensor \eqref{embed}. The main observation is that the Chern-Simons terms \eqref{CSterm} and \eqref{CSg} also match.

\subsubsection{Potential}\label{potsec}
The only term in the Lagrangian \eqref{3dgauged} left to compute is the potential $\theletterwhichIwilluseinthefollowingforthescalarpotential$. Note that the comparison of the other terms in \eqref{3dgauged} with those of \eqref{3dtheory} has fixed all freedom in field redefinitions or identification of coupling constants. So comparison of the potentials will be even more non-trivial. 

Computing the scalar potential in gauged supergravity can often be a daunting task, due to its complicated nature. Here however we will use an observation that for a large class of $D=3$ $N=4$ gauged supergravities the potential is completely determined in terms of a single superpotential function. We first discuss this result, that can be of interest in a wider context, and then apply it to compute the potential of the theory we are considering.

\vspace{0.2cm}

\noindent{\bf{Superpotential for degenerate $\susyN=4$ theories}}

\noindent The general formula for the scalar potential of three-dimensional supergravity 
was derived in \cite{deWit:2003ja} and for $\susyN=4$ it reads
\begin{eqnarray}
\theletterwhichIwilluseinthefollowingforthescalarpotential
=\frac{1}{4}\scalmet^{\iscala\iscalb}D_\iscala \Apot_1^{\ifoura\ifourb}D_\iscalb \Apot_1^{\ifoura\ifourb}-2\Apot_1^{\ifoura\ifourb}\Apot_1^{\ifoura\ifourb}
+\scalmet^{\iscala\iscalb}\Tten_\iscala{}^{\rproj{\ifoura\ifourb}}\Tten_\iscalb{}^{\rproj{\ifoura\ifourb}}\;,
\label{potential}
\end{eqnarray}
where
\begin{equation}
\Apot_{1}^{\ifoura\ifourb}\equiv -2 \Tten^{\rproj{\ifoura\ifourc},\rproj{\ifourc\ifourb}}+\frac{1}{3}\delta^{\ifoura\ifourb}\Tten^{\rproj{\ifourc\ifourd},\rproj{\ifourc\ifourd}}\;.
\end{equation}

It is determined in terms of the metric of the scalar target space $\scalmet_{\iscala\iscalb}$ and invariant under scalar field redefinitions and R-symmetry transformations through the appearance of the mixed diffeomorphism and SO(4)$_\labelsofR$ covariant derivative $D_\iscala$. The last ingredient is the T-tensor, which is defined in terms of the embedding tensor and the $\cosv$ matrices that appear in the transformations of the fermions \cite{deWit:2003ja}:
\begin{equation}
\Tten_{\igena,\igenb}=\cosv^{\igenc}{}_{\igena}\cosv^{\igend}{}_{\igenb}\embed_{\igenc,\igend}
\;.
\end{equation}
As before the $\ifoura,\ifourb$ indices are fundamental SO(4) indices, furthermore  with $\rproj{\ifoura\ifourb}$ we denote the projection along the adjoint representation of the R-symmetry group SO(4)$_\labelsofR$.

As shown in \cite{deWit:2003ja}, the case $\susyN=4$ is somewhat special as supersymmetry requires the scalar target space to be a direct product of two quaternionic manifolds of dimensions $d_+=4n_+$ and $d_-=4n_-$. The SO(4)$_\labelsofR$ splits as SO(3)$_+\times$SO(3)$_-$ where the first/second factor only acts non-trivially on the first/second factor of the scalar manifold respectively.
There is a degenerate case where one of the two quaternionic manifolds, say the second, is just a point: $n_-=0$, which is the case for our target space (\ref{so4444}). In this case the theory simplifies considerably and, as we will now show, the potential \eqref{potential} can be written in terms of a real superpotential, provided the SO(3)$_-$ factor of the R-symmetry remains ungauged.

It was derived in \cite{deWit:2003ja} that in this degenerate $\susyN=4$ case supersymmetry requires the pure R-symmetry components of the T-tensor to be a singlet under that symmetry:
\begin{equation}
\Tten^{\rproj{\ifoura\ifourb},\rproj{\ifourc\ifourd}}=\acronymforthethreeDsuperpotential\, \Pp^{\ifoura\ifourb,\ifourc\ifourd}\,, \qquad \Pp^{\ifoura\ifourb\ifourc\ifourd}\equiv\frac{1}{4}\left(\delta^{\ifoura\ifourc}\delta^{\ifourb\ifourd}
-\delta^{\ifourb\ifourc}\delta^{\ifoura\ifourd}+\epsilon^{\ifoura\ifourb\ifourc\ifourd}\right)\,.\label{specform}
\end{equation}
Now note that this restrictive form also determines the mixed T-tensor components appearing in \eqref{potential}. This follows from the general relation \cite{deWit:2003ja}   
\begin{equation}\label{Trel}
D_\iscala \Tten^{\rproj{\ifoura\ifourb},\rproj{\ifourc\ifourd}}=\frac{1}{2}\fk^{\rproj{\ifoura\ifourb}}_{\iscala\iscalb}\Tten^{\rproj{\ifourc\ifourd},\iscalb}+
\frac{1}{2}\fk^{\rproj{\ifourc\ifourd}}_{\iscala\iscalb}\Tten^{\rproj{\ifoura\ifourb},\iscalb}
\;.
\end{equation}
On the right hand side the SO(4)$_\labelsofR$ adjoint-valued K\"ahler forms on the scalar manifold appear. In the case we are discussing, these K\"ahler forms are only non-trivial on one factor and it is convenient to define
\begin{equation}
\fk_+^{\ifoura\ifourb}\equiv\Pp^{\ifoura\ifourb,\ifourc\ifourd}\fk^{\ifourc\ifourd}
\;.
\end{equation}
Here the $\fk_+^{\isota 4}$, $\isota=1,2,3$, are manifestly SO(3)$_+$ covariant and form a quaternionic algebra. In case the SO(3)$_-$ factor is not gauged, the T-tensor will have no components along it, i.e. $\Tten^{\iscala,\rproj{\ifoura\ifourb}}=\Pp^{\ifoura\ifourb\ifourc\ifourd}\Tten^{\iscala,\rproj{\ifourc\ifourd}}$. This constraint together with \eqref{specform} implies there is a unique solution to \eqref{Trel}, which after some manipulations using the quaternionic algebra can be written as
\begin{equation}\label{Tresult}
\Tten^{\rproj{\ifoura\ifourb}}{}_{\iscala}=\frac{1}{4}\fk^{\ifoura\ifourb}_{+\,\iscala\iscalb}
\partial^\iscalb \acronymforthethreeDsuperpotential\,. 
\end{equation}
One can now simply plug the expressions \eqref{specform} and \eqref{Tresult} into the formula \eqref{potential} and find that the invariant $\acronymforthethreeDsuperpotential$ is nothing but a real superpotential
\begin{equation}
\theletterwhichIwilluseinthefollowingforthescalarpotential
=\scalmet^{\iscala\iscalb}\partial_\iscala \acronymforthethreeDsuperpotential\partial_\iscalb \acronymforthethreeDsuperpotential-2\acronymforthethreeDsuperpotential^2\;.\label{pot}
\end{equation}

\vspace{0.2cm}

\noindent{\bf{Computing the potential}}

\noindent
We can now use the expression of the potential in terms of the superpotential \eqref{pot} to explicitly find the potential of our $D=3$ gauged supergravity. To find the superpotential we compute the T-tensor via \eqref{embed} and (\ref{cosstuff}), and find that the only non-vanishing components (up to index symmetries) are
\begin{eqnarray}
\Tten_{\ip\jp,\kp\lp}&=&\frac{\gprime}{2}\epsilon_{\ifoura\ifourb\ifourc\ifourd}\det\sqrt{\Tmat}\;,\\
\Tten_{\ip\jp,\kp\lm}&=&-\frac{\popeg}{2}\left(\Tmat_{\ifoura\ifourc}\delta_{\ifourb\ifourd}-\Tmat_{\ifourb\ifourc}\delta_{\ifoura\ifourd}\right)\;.
\end{eqnarray}
The second step is to project these components along the R-symmetry. As discussed at the beginning of this section we identified the SO(3)$_\labelsota$ as the R-symmetry factor acting on the scalar manifold. Its generators are embedded in so(4,4) as follows:
\begin{equation}\label{Rgen}
\sotagen^{\ifoura\ifourb}=\frac{1}{2}\Pp^{\ifoura\ifourb\ifourc\ifourd}\left(\torgen_+^{\ifourc\ifourd}+\torgen_-^{\ifourc\ifourd}+\dsofgen^{\ifourc\ifourd}\right)\,.
\end{equation}
This implies that for a generic SO(4,4) adjoint valued tensor $\ften_{\ieighta\ieightb}$ the projection is defined as:
\begin{eqnarray}
\ften_{\rproj{\ifoura\ifourb}}&\equiv&\frac{1}{2}\Pp^{\ifoura\ifourb\ifourc\ifourd}\left(\omega_{\kp\lp}+\omega_{\km\lm}+
\omega_{\kp\lm}+\omega_{\km\lp}\right)\\
&\Rightarrow& \ften_{\ieighta\ieightb}\soegen^{\ieighta\ieightb}=\ften_{\rproj{\ifoura\ifourb}}\sotagen^{\ifoura\ifourb}+
\mbox{non-R-symmetry generators}\,.
\end{eqnarray}  
It is then a matter of algebra to compute that
\begin{equation}
\Tten_{\rproj{\ifoura\ifourb},\rproj{\ifourc\ifourd}}=\frac{1}{2}\left(\gprime\det\sqrt{\Tmat}-\popeg\Tr\,\Tmat\right)\Pp^{\ifoura\ifourb\ifourc\ifourd}
\;.
\end{equation}
First of all it is important to note that the T-tensor is of the form \eqref{specform}, which implies that our choice \eqref{embed} for the embedding tensor is compatible with supersymmetry! Furthermore it allows us to read off the superpotential
\begin{equation}
\acronymforthethreeDsuperpotential=\frac{1}{2}\left(\gprime\det\sqrt{\Tmat}-\popeg\Tr\,\Tmat\right)
\;.
\end{equation}
Together with the scalar metric \eqref{kin} we can then finally compute the potential:
\begin{eqnarray}
\theletterwhichIwilluseinthefollowingforthescalarpotential
=\frac{1}{2}\left(\gprime^2\det\Tmat+2\popeg^2\Tr\,\Tmat^2-\popeg^2(\Tr\,\Tmat)^2\right)
\label{pot2}
\end{eqnarray}
It is very gratifying to see that indeed this potential derived by imposing supersymmetry on the
lower-dimensional Lagrangian, matches the potential resulting from the compactification \eqref{potred} in all detail.

\section{Discussion}

In this paper we have shown that the $S^3$ reduction of $D=6$ $\susyN=(1,0)$ supergravity coupled to a single chiral tensor-multiplet
gives rise to a Lagrangian (\ref{3dtheory}) that falls within the class of $N=4$ gauged supergravities in three dimensions. 
This defines a special case of the general family of reductions
constructed in \cite{Cvetic:2000dm}, which allows for a supersymmetric effective theory and a supersymmetric AdS$_3$ vacuum.
Around this vacuum, the 10 scalar degrees of freedom split into a singlet of conformal dimension $\Delta=4$ and 9 massless
scalars in the irreducible $(1,1)$ representation of the SO$(4)$ gauge group.

Since much work has been devoted to various reductions from six down to three dimensions, let us comment on
the relation of the present model to other known compactifications. An SU$(2)$ group manifold reduction of (\ref{Lag6}) has
been studied in \cite{Gava:2010vz}. Such a reduction is automatically consistent by symmetry and in this case
(in the absence of $D=6$ vector multiplets) induces a three-dimensional theory with SO$(3)$ gauge group. The resulting theory is
described by the truncation of (\ref{3dtheory}) to singlets under one SO$(3)$ factor of the gauge group. This corresponds
to the reduction of the scalar target space (\ref{so4444}) to SO$(4,1)/$SO$(4)$ and the potential (\ref{pot3}) to
\begin{eqnarray}
\theletterwhichIwilluseinthefollowingforthescalarpotential &=&
2\,g_0^2 \left(e^{4\sigma}- 2\,e^{2\sigma} \right)
\;,
\label{pots1}
\end{eqnarray}
for the surviving scalar field of $\Delta=4$.
A different SU$(2)$ group manifold reduction has been worked out in~\cite{Lu:2002uw,Lu:2003yt}. Here, the starting point is the
pure (chiral) $N=(1,0)$ theory and the volume mode $\phi$ of the sphere is part of the three-dimensional scalar sector.
In this case, the potential for the volume mode is of the form\footnote{ Normalized with respect to the scalar kinetic term as was done for (\ref{pots1}).}
\begin{eqnarray}
\theletterwhichIwilluseinthefollowingforthescalarpotential &=&
 2 g_0^2 \left(2\,e^{2\sqrt{3}\,\phi} -3\,e^{4\phi/\sqrt{3}} \right)
 \;,
\label{V32}
\end{eqnarray}

It is easy to check that this potential equally describes a scalar of conformal dimension $\Delta=4$,
but with a profile different from (\ref{pots1}) beyond the quadratic approximation.
The resulting $D=3$ theory thus cannot be obtained as a truncation from (\ref{3dtheory}), but rather
corresponds to a different gauging of SO$(3)$ within the isometries of the relevant scalar target space 
SO$(4,3)/($SO$(4)\times$SO$(3)$)\,.\footnote{
Such gaugings have also been studied in \cite{Karndumri:2012qn}
however with an ansatz that only captures one of the two terms in (\ref{V32}).}

An interesting generalization of the present construction would be a possible embedding
into a larger consistent truncation preserving more supersymmetries. Indeed, the 
three-dimensional theory (\ref{3dtheory}) has a natural embedding into the $N=8$
gauged supergravity with coset space $SO(8,4)/(SO(8)\times SO(4))$
and gauge group $SO(4)$ embedded in a diagonal way, as constructed in~\cite{Nicolai:2001ac,Nicolai:2003ux},
which reproduces the scalar potential (\ref{pots1}) upon proper truncation.
In six dimensions, this should correspond to an embedding of (\ref{Lag6}) into the half-maximal 
$N=(1,1)$ theory. Specifically, the 16 extra fields in the three-dimensional scalar target space
should have a higher-dimensional origin among the internal components of the 4 additional vector fields
of the $D=6$ $N=(1,1)$ supergravity multiplet. 
Even more challenging would be the extension of the present construction to a half-maximal 
reduction within the chiral $D=6$ $N=(2,0)$ theory into which (\ref{Lag6}) can be embedded upon adding 4 additional
chiral tensor multiplets. The complete spectrum of its AdS$_3 \times S^3$ compactification has been obtained in \cite{Deger:1998nm}. 
Even though there is a unique three-dimensional $N=8$ supergravity which reproduces precisely the linearized spectrum of this compactification~\cite{Nicolai:2003ux}, it somewhat mysteriously fails to reproduce the correct profile (\ref{V32}) of the $S^3$ volume mode.
A possible consistent truncation of the $N=(2,0)$ theory preserving all supersymmetries thus remains an open problem
and may require additional matter couplings in six dimensions.

Let us finally mention that more recently the construction of duality covariant formulations of higher-dimensional supergravities
allow to reconsider and address the question of consistent truncations in a more abstract 
and very powerful framework~\cite{Aldazabal:2013mya,Hohm:2013pua,Lee:2014mla,HSKK} in the spirit of the original work~\cite{deWit:1986iy}.

\section*{Acknowledgments}
The authors would like to thank E. Sezgin for helpful discussions. NSD and \"OS are partially supported by TUBITAK grant 113F034. 
DVdB is partially supported by TUBITAK grant 113F164 and Bo\u{g}azi\c{c}i University Research Fund BAP grant 13B03SUP7.

\appendix

\section*{Appendix}
\section{Normalization of the scalar kinetic term}\label{norm}
The coupling of $D=3$ non-linear sigma models to gravity in a supersymmetric fashion implies a 
precise normalization of the scalar metric \cite{deWit:1992up, deWit:2003ja}. It manifests itself as the relation (\ref{normrel}) 
between the SO($\susyN$) R-symmetry valued connection and K\"ahler form. As the right-hand side of this relation depends explicitly 
on the scalar metric whereas the left-hand side does not, it fixes a preferred normalization.

We will now show how this relation is indeed satisfied in our model with the normalization as in section \ref{gssection}. In our case the scalar manifold is the coset SO(4,4)$/$(SO(4)$\times$SO(4)) and both the K\"ahler forms and the R-symmetry connection are fully determined by the symmetries. We can construct a coset representative as follows
\begin{equation}
\cosrep=e^{\escalars_{\ifoura\ifourb}\torgen_+^{\ifoura\ifourb}}e^{\scalars_{\ifourc\ifourd}
\qgen^{\ifourc\ifourd}}=\begin{pmatrix}
\sqrt{\Tmat^{-1}}&-2\escalars\sqrt{\Tmat}\\
0&\sqrt{\Tmat}
\end{pmatrix}\,.
\end{equation}
One can then compute that
\begin{equation}
\cosrep^{-1}d\cosrep=\frac{1}{2}\left(\sqrt{\Tmat^{-1}}d\sqrt{\Tmat}\right)_{\ifoura\ifourb}\left(\qgen^{\ifoura\ifourb}
-\dsofgen^{\ifoura\ifourb}\right)+
\left(\sqrt{\Tmat}d\escalars\sqrt{\Tmat}\right)_{\ifoura\ifourb}\torgen_+^{\ifoura\ifourb}\,.
\end{equation}
The vielbein $e$ and the R-symmetry valued part of the spin-connection $\cosq_\labelsota$ are defined as
\begin{equation}
\cosrep^{-1}d\cosrep=\cosp^{\ifoura\ifourb}\Ync^{\ifoura\ifourb}-\frac{1}{2}\cosq^{\ifoura\ifourb}_\labelsota\sotagen^{\ifoura\ifourb}+\ldots\,,
\end{equation}
where the R-symmetry generators $\sotagen^{\ifoura\ifourb}$ were defined in \eqref{Rgen}, we collected the non-compact generators of so(4,4) as follows
\begin{equation}
\Ync^{\ifoura\ifourb}=\frac{1}{2}\left(\torgen_+^{\ifoura\ifourb}-\torgen_-^{\ifoura\ifourb}+\qgen^{\ifoura\ifourb}\right)\,,
\end{equation}
and we omitted terms proportional to other generators.
It follows from these definitions that
\begin{eqnarray}
\cosp^{\ifoura\ifourb}&=&\left(\sqrt{\Tmat^{-1}}d\sqrt{\Tmat}\right)_{(\ifoura\ifourb)}+
\left(\sqrt{\Tmat}d\escalars\sqrt{\Tmat}\right)_{\ifoura\ifourb}\,,\\
\cosq^{\ifoura\ifourb}_\labelsota&=&\Pp^{\ifoura\ifourb\ifourc\ifourd}
\left(\left(\sqrt{\Tmat^{-1}}d\sqrt{\Tmat}\right)_{\ifourc\ifourd}+
\left(\sqrt{\Tmat}d\escalars\sqrt{\Tmat}\right)_{\ifourc\ifourd}\right)\,.
\end{eqnarray}

The complex structures can be directly related to the representation of the non-compact generators under the R-symmetry:
\begin{equation}
[\sotagen^{\ifoura\ifourb},\Ync^{\ifourc\ifourd}]\equiv\frac{1}{2}\Gam^{\ifoura\ifourb\, \ifourc\ifourd}{}_{\ifoure\ifourf}\Ync^{\ifoure\ifourf}\,, \qquad \Gam^{\ifoura\ifourb\, \ifourc\ifourd}{}_{\ifoure\ifourf}= 4\delta^{\ifourc\ifoure}\Pp^{\ifoura\ifourb\ifourd\ifourf}\,.
\end{equation}
The K\"ahler forms are then \cite{deWit:2003ja}:
\begin{eqnarray}
\fk^{\ifoura\ifourb}&=& -\Gam^{\ifoura\ifourb}{}_{\ifourc\ifourd\,\ifoure\ifourf}\cosp^{\ifourc\ifourd}\wedge\cosp^{\ifoure\ifourf}\,.
\end{eqnarray}
It is now a matter of (somewhat tedious) algebra to verify that indeed the relation \eqref{normrel} is satisfied.


\providecommand{\href}[2]{#2}\begingroup\raggedright\endgroup

\end{document}